\documentstyle[aps,amssymb,12pt]{revtex}

\begin{document}
\author{Giuseppe Castagnoli\thanks{%
Information Technology Dept., Elsag Bailey, 16154 Genova, Italy}}
\title{{\Huge Quantum Computation Based on Retarded and Advanced Propagation } }
\maketitle
\date{November 19, 1997}

\begin{abstract}
Computation is currently seen as a forward propagator that evolves (retards)
a completely defined initial vector into a corresponding final vector.
Initial and final vectors map the (logical) input and output of a reversible
Boolean network respectively, whereas forward propagation maps a one-way
propagation of logical implication, from input to output.

Conversely, hard $-$ NP-complete $-$ problems are characterized by a two-way
propagation of logical implication from input to output and vice versa,
given that both are partly defined from the beginning. Logical implication
can be propagated forward and backward in a computation by constructing the
gate array corresponding to the entire reversible Boolean network and
constraining output bits as well as input bits. The possibility of modeling
the physical process undergone by such a network by using a retarded and
advanced in time propagation scheme is investigated.

\end{abstract}

\section{The Notion of Computation Reversibility}

This work deals with the possibility of deepening the notion of computation
reversibility in the quantum framework. This might provide a richer form of
computation.

The discovery that computation can be reversible in principle and the
development of the thought model of a classical reversible computer, became
cornerstones of research on the physical basis of computation [1, 2, 3, 4].
Once found to be reversible, computation could be moved into the quantum
framework [5, 6, 7, 8, 9].

The finding that the deeper the physical basis the richer is the form of
computation, has been a remarkable result:

\begin{itemize}
\item  a quantum register undergoing a unitary evolution can perform in
quantum superposition (or parallelism) a number of computations which grows
exponentially with register size,

\item  superposition elements interfere and the successive measurement
brings about a collective property of those elements in the episystem,

\item  this enables faster than classical computation in some special
instances [7, 10, 11, 12, 13, 14].
\end{itemize}

Although essential to bring computation into the quantum domain, the notion
of computation reversibility entered the quantum framework unaltered. This,
we argue, might constitute an undue restriction.

As is well known, the notion of reversibility, thus of the absence of a
privileged direction of causality, can undergo a quantum change. Models that
postulate the non-redundant coexistence of a retarded and an advanced
propagation in the quantum framework have been formulated [15, 16, 17, 18,
19], although mostly outside a computational context.

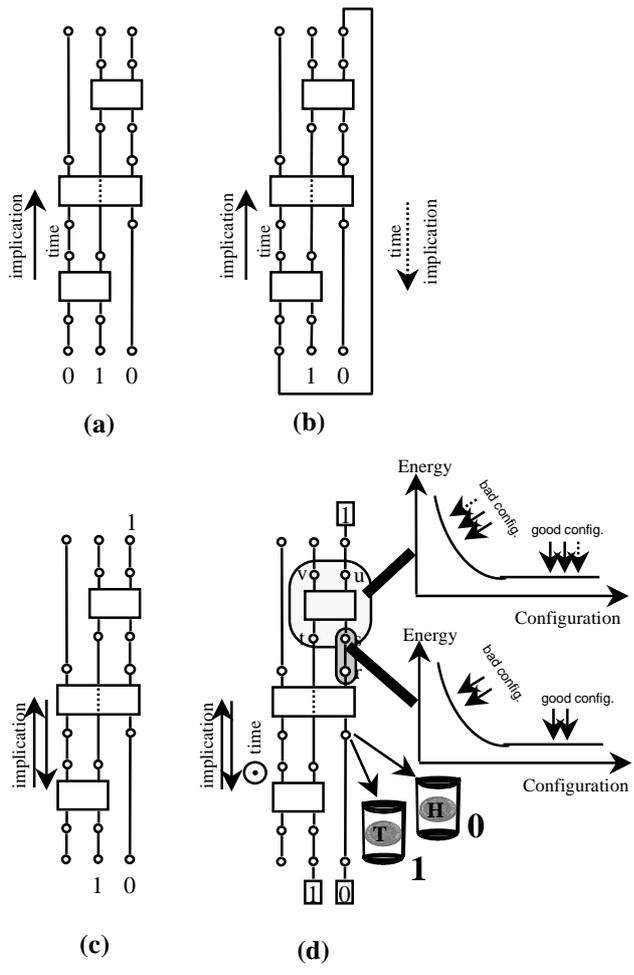
\begin{figure}[tbp]
\caption{}
\label{fig:1}
\end{figure}

Let us examine the notion of computation under a broader perspective,
keeping in mind that it can be mapped on either a one or a two-way
propagation. Currently, quantum computation is seen as a propagator that
evolves (retards) a completely defined initial vector into a final vector,
which is therefore a function of the initial vector. One-way physical
propagation maps the one-way propagation of logical implication from the
inputs to the outputs of a reversible Boolean network (Fig.~\ref{fig:1}a).
Such a network appears in the space-time diagram of a ``sequential
computation'' process: network nodes are the register qubits {\em before}
and {\em after} each transformation, Boolean gates and wires represent the
transformations.

The existence of a privileged input-output direction (of one-way
propagation) in sequential computation sets some constraints on the
underlying Boolean network. A\ Boolean constant must be preassigned to any
input node, whereas none is allowed on output nodes. An output cannot be
connected to an upstream input which participates in the logical
determination of that same output (Fig.~\ref{fig:1}b; ignore dotted arrow).
The fact that sequential computation is suspected to be unable to solve
NP-complete problems in polynomial time [20] might be related to the
presence of such constraints.

Conversely, Boolean networks do appear in problem solving in a less
constrained way. An example is the problem of finding the satisfiability of
a {\em generic} Boolean network. Fig.~\ref{fig:1}c exemplifies the standard
way of stating this problem for a reversible network. Out of habit, the two
extremities of the network are still called input and output, although there
is no longer a privileged input-output direction. Both the input and the
output must be {\em partly} constrained: constant Boolean values must be
assigned to part of the input nodes, another part must be left free. The
same applies to the output. The Boolean constants applied to both the input
and the output propagate a {\em conditional} {\em logical implication},
respectively in the input-output and output-input directions (of course the
logical implication, say, in the input-output direction is conditioned by
the possible Boolean assignments of the unconstrained part of the input, and
similarly for the other direction). The coexistence of two directions of
conditional implication makes the problem of determining network
satisfiability hard, namely NP-complete.

The case of the Boolean network in Fig.~\ref{fig:1}b (ignore the time axis)
is similar. The output-input connection introduces another form of
coexistence of a forward and backward conditional implication (along a
loop), which makes the network satisfiability test equally hard.

It is physically quite possible to propagate a conditional logical
implication both forward and backward along a reversible Boolean network, by
laying out the entire network spatially and by constraining part of the
inputs as well as part of the outputs (Fig.~\ref{fig:1}d). Thus network
nodes $r,s$, ... become {\em coexisting} qubits $\left| \chi
_{1}\right\rangle _{r}$, $\left| \chi _{2}\right\rangle _{s}$, ..., (with $%
\chi _{i}=0,1$) as time is now perpendicular to the network layout.

We will develop a thought model of this network by applying a sort of
divide-and-rule strategy, meant to take advantage of a quantum symmetry
related to particle statistics\footnote{%
So far quantum computation has used special features like superposition and
multiparticle interference, certainly not particle statistics. The
possibility of exploiting particle statistics is investigated in this work.}.

Conceptually, gates, with their input and output qubits, are first thought
of as being ``divided'', all the constraints imposed on them through the
wires are removed. Each independent (unwired) gate is implemented by a
quantum transposition of simulated annealing---namely ground state
computation. At equilibrium, each gate sits in a degenerate ground state.
This can host any linear combination of those gate eigenstates---each a
tensor product of the gate qubits eigenstates---which satisfy the gate
Boolean relation. Disregarding wires, all gates would relax toward ground in
parallel and independently of each other. This, at equilibrium, would
produce a network state consisting of a superposition of tensor products
which satisfy the Boolean relations imposed by the gates, while those
imposed by the wires are ignored.

Wires are restored by using a special symmetry induced by particle
statistics. With no loss of generality, each wire is assumed to comprise a
NOT\ function. As such, it will be called link. The truth table of link,
say, $r,s$ is thus $
\begin{array}{ll}
r & s \\ 
0 & 1 \\ 
1 & 0
\end{array}
.$ Such a relation between the eigenvalues of any two interlinked qubits $%
r,s $, will be imposed by a special symmetry $A_{rs}$, related to particle
statistics [defined by $A_{rs}\left| \chi \right\rangle _{r}\left| 1-\chi
\right\rangle _{s}=\left| \chi \right\rangle _{r}\left| 1-\chi \right\rangle
_{s},$ $A_{rs}\left| \chi \right\rangle _{r}\left| \chi \right\rangle
_{s}=0, $ with $\chi =0,1$---see Section II.C]. Such symmetries, or their
product for all links $\tilde{A}=\Pi _{r,s}A_{rs}$ (the $A_{rs}$ pairwise
commute), ``rule'' by restoring consistent wiring and network completeness.

This kind of wiring is liable of two interpretations, one holding in the
conventional one-way propagation model, the other in a two-way propagation
model (by ``propagation'', when not explicitly referred to conditional
logical implication, we will mean the physical process). Both will be
pursued throughout this work.

In one-way propagation, one obtains a quantum annealing computation over a
Boolean network. However, no advantage will be found over classical
annealing (disregarding quantum tunnelling, not addressed in this work).

In two-way propagation, it is as if $\tilde{A}$ induced a watchdog effect on
the overall network state $\left| \psi \left( t\right) \right\rangle $
generated by the relaxation process of the {\em independent} gates. By the
way, we should note that $\left| \psi \left( t\right) \right\rangle $ is not
necessarily a pure state, we will use the method of random phases.
Symmetrization under $\tilde{A}$ would operate as projection, canceling so
to speak the amplitudes of those network eigenstates which would violate $%
\tilde{A}$, and reinforcing the others through renormalization. This would
yield (roughly) to a population of gates relaxing independently of each
other, without ever incurring frustration, with an essential advantage over
classical annealing\footnote{%
To some extent, this is similar to present-day quantum error correction,
where the onset of errors and decoherence is countered by repeated
projection of a redundant computation state onto some symmetric subspace
[21, 22, 23, 24].}.

In this scheme, any particle statistics symmetry $S$ (for example, singlet
or triplet state symmetry) must be seen as a projector $S$ applying to the
evolving state for all times $t$: $\forall t:S\left| \psi \left( t\right)
\right\rangle =\left| \psi \left( t\right) \right\rangle $. In this view, $S$
does not need to be an initial condition conserved as a constant of motion.
Eventually, $S$ commutes with the propagator of $\left| \psi \left( 0\right)
\right\rangle $ ; but this can be so in a reversed perspective, where the
propagator could be {\em forged} by projection.

Considering a particle statistics symmetry $S$ as the result of projection,
is consistent with the usual interpretation where symmetry is seen as a
conserved initial condition. However, the two interpretations are not
necessarily equivalent: in the current computation context, the projection
view appears to favour a two-way propagation model. We mean those
interpretations of quantum mechanics which postulate that propagations are
driven by {\em both} their initial and final conditions$^{\left[
15,16,17,18,19,33\right] }$.

In current assumptions we are dealing with a set of independently relaxing
gates, whose overall state, for all times $t$, is projected on the Hilbert
subspace which satisfies symmetry $\tilde{A}$: propagation would thus be
``driven'' by relaxation and ``forged'' by projection.

Let us consider first the usual one-way propagation model. Here propagations
are modeled by calculus, which is of course one-way. Any infinitesimal
relaxation should be submitted to projection, and so on repeatedly. However,
this yields to a Zeno effect that ``freezes'' the propagation in its initial
state: $\forall t:\left| \psi \left( t\right) \right\rangle =\left| \psi
\left( 0\right) \right\rangle $ (Section II.D).

It might be argued that one-way calculus is not fit to model a propagation
``forged'' by projection. Intuitively, the nature of projection appears to
be discontinuous; even if projection is applied at all times $t$, this does
not necessarily mean that it can be used in a differential way. As a matter
of fact, the Zeno effect disappears when this projection-forged propagation
is modeled by the two-way propagation scheme developed in [18].

Here the propagation of $\left| \psi \left( t\right) \right\rangle $ turns
out to be driven by the final condition that projection at all times $t$
yields a state that {\em simultaneously} (under a system of {\em %
mathematically} {\em simultaneous} equations) approaches ground, according
to the independent gate relaxation process, and satisfies $\tilde{A}$.

Interestingly, projection on a predetermined outcome implies an elusive
chronology violation. At any time $t_{1}$ the propagation has to choose
between two possible projection outcomes, one satisfying, the other not
satisfying $\tilde{A}$. Satisfaction of $\tilde{A}$ is the condition of that
choice. The propagation meets this condition {\em immediately} {\em after}
the choice:\ the choice should be before projection (at time $t_{1}$), and
the condition is met after projection (at time $t_{2}$, reasonably $%
t_{2}>t_{1}$). Chronology violation is elusive since it can be confined
within a time interval $\Delta t=t_{2}-t_{1}$ as short as desired. It can
be, so to speak, ``safely'' confined within a time interval $\Delta t$ such
as $\Delta t\Delta E\cong 
\rlap{\protect\rule[1.1ex]{.325em}{.1ex}}h%
$, where the ordering of time and causality becomes fuzzy. However, the
consequences of choices taken in $\Delta t$ can naturally be ``long range''.
For the sake of clarity, we should note that chronology violation would
become completely evident if the vector to be projected on a {\em %
predetermined} outcome were discretely different from it. A condition set in
the future would determine a choice in the past. It becomes elusive when
projection is performed for all $t$.

Section II provides a thought model based on the above notions. It is not an
implementation model; some Hamiltonians are represented by Hermitean
matrices. However, it appears to be plausible in principle. If
implementable, it could speed up the solution of an NP-complete problem with
respect to known classical computation.

An analogy can be observed between this work and D. Deutsch's work on closed
time-like lines [26]. Their relation to general relativity is irrelevant in
this context $-$ what is relevant is the presence of loops of conditional
implication. In Fig.1b, forward and backward conditional implications
coexist on the two halves of a loop which (in Deutsch's model) becomes a
closed time-like line in a chronology violating region (consider now the
dotted arrow). Sequential computation circulates along this line. This
enables, without necessarily implying it, faster than classical computation
(the problem was finding the fixed points of a hard to invert function).

\section{Thought Model}

\subsection{Classical Simulated Annealing}

With a view to introducing quantum annealing, we shall review a simple
application of {\em classical} simulated annealing $-$ also called ground
state computation $-$ to the satisfiability problem. Network nodes (Fig. 1d)
can be seen as cups, each containing a coin in either head or tail state
(say Boolean 0 or 1). Computation follows an {\em if/then} process:

\begin{enumerate}
\item[(1)]  does the coin configuration satisfy all the local relations
imposed by gates and links?

\item[(2)]  If YES, stop: a solution has been found and the network is
satisfiable.

\item[(3)]  If NOT, ``shake'' (randomly flip coins) and go to (1).
\end{enumerate}

\noindent Network unsatisfiability is ascertained with any desired
confidence level by looping for a sufficiently large number of times,
without ever finding a solution.

In order to speed up computation, each network element (N.E.) should be
viewed as a mechanism whereby all states which satisfy the N.E. belong to a
degenerate ground state, while the energy level of the others is discretely
above ground (Fig.~\ref{fig:1}d). This feature $-$ during ``shaking'' $-$
provides an energy gradient, orienting the evolution towards ground state.
Dumping is of course needed. A heat bath of suitable temperature can provide
for both dumping and shaking.

Shaking is required because this kind of computation is subject to
``frustration''. A coin configuration can satisfy most N.E. and still be far
from any solution. The process, so to speak, should be undone. This boils
down to the presence of {\em relative} energy minima $-$ ``shaking'' is
needed to avoid state trapping in such minima. Let N be the number of
network nodes. The number of relative minima and relaxation time $-$
expected time to reach the absolute minimum $-$ is exponential in N,
according to today's knowledge by any classical simulated annealing
strategy. For a description of efficient strategies, see [27].

\subsection{Quantum Annealing $-$ Gates}

In quantum annealing, the Hamiltonian of the Boolean gate operating on
qubits, say, $s,t,u,v$ (Fig.~\ref{fig:1}d), has 16 eigenstates, namely all
possible tensor products of the gate qubit eigenstates. Those tensor
products which satisfy the gate belong to a degenerate ground state, while
those which do not satisfy it are metastable states belonging to discretely
higher, say by more than $\Delta E,$ energy levels. An external input or
output with a preassigned Boolean value becomes a {\em one-qubit gate}, and
that value will correspond to the ground state of the gate Hamiltonian (Fig.~%
\ref{fig:1}d).

Given a heat bath temperature $T$ such that $0<kT<<\Delta E$, in a
semiclassical picture the probability $p$ of finding (as the result of a
measurement) the gate in ground state by the time $t$ follows asymptotically
the law 
\begin{equation}
p=1-e^{-\sigma t},
\end{equation}

\noindent where $\sigma >0.$ Of course this exponential law holds when
relaxation has reached a constant rate. Model Hamiltonians of elementary
Boolean gates with {\em coexisting} inputs and outputs can be found in [28].
Of course such gates differ from the gates used in the sequential approach,
where inputs and outputs are successive states of the same register [6, 7,
29, 30, 31, 32].

Links could be implemented accordingly. Then, link-gate interaction
Hamiltonians should be added. The number of local minima and computation
time would be likely to remain exponential in $N$ (the benefit deriving from
quantum tunnelling through local minima is not considered in this work).
Links will be dealt with in a different way.

\subsection{Quantum Annealing and Particle Statistics $-$ Links}

The correlation between simultaneous qubit eigenvalues imposed by links will
rely on a special particle statistics symmetry $-$ seen as a projection
operator (Section I). The (thought) physical situation giving origin to such
symmetry (one for each link) is still obtained by means of a relaxation
process.

Without any restriction, a network topology is adopted where nodes come in
interlinked pairs (Fig.~\ref{fig:1}). The ``divide'' rule is still applied.
By ``turning off'' gate Hamiltonians, links are, so to speak, ``ungated''.
This leaves us with a set of independent links, which comprise all the nodes
of the network.

For the sake of visualization, each link node $r,s$ (Fig. \ref{fig:1}d) can
be seen as a site of a spatial lattice. These sites are labeled by $\lambda
=r,s$. Each site hosts exactly one spin $\frac{1}{2}$ particle, with its $%
\sigma _{z}$ spin component either down $\left( \chi =0\right) $ or up $%
\left( \chi =1\right) $. Therefore qubit $\left| \chi \right\rangle
_{\lambda }$ reads: the $\sigma _{z}$ component of the particle hosted by
site $\lambda $ is $\chi $. However, we should like to keep this model
abstract: $\lambda $ and $\chi $ can be any two binary compatible attributes
(dynamical variables) of a fermion.

In a particle label representation, qubit $\left| \chi \right\rangle
_{\lambda }$ is written $\left| \chi \right\rangle _{1}$ $\left| \lambda
\right\rangle _{1}$; for example $\left| 0\right\rangle _{r}$ becomes $%
\left| 0\right\rangle _{1}\left| r\right\rangle _{1}$. By this we mean that
the $\chi $ attribute of particle 1 has eigenvalue 0 and that its $\lambda $
attribute has eigenvalue $r$. Or, since we are dealing with two identical
particles, particle 1 is interchangeable with particle 2; in this case $%
\left| 0\right\rangle _{r}$ becomes $\left| 0\right\rangle _{2}\left|
r\right\rangle _{2}$. Therefore, qubit representation will be the outcome of
a finer identical particle representation.

The lattice $L$ of network nodes (or qubits), with exactly one fermion per
node (one independent $\chi $ attribute per qubit), is created as follows.
Gate Hamiltonians are still ``turned off''. Link Hamiltonians, thus
completely independent of each other, are ``turned on''. Then links are
allowed to relax on their ground states.

The Hamiltonian $H_{12}^{\lambda }$ of link $r,s$ is defined as follows. It
operates on the spatial coordinates of fermions 1 and 2 $-$ {\em not} on
their spin. Therefore $H_{12}^{\lambda }$ operates on the Hilbert space:

\[
H_{\lambda }=span\left\{ \left| r\right\rangle _{1}\left| r\right\rangle
_{2},\left| s\right\rangle _{1}\left| s\right\rangle _{2},\frac{1}{\sqrt{2}}%
\left( \left| r\right\rangle _{1}\left| s\right\rangle _{2}-\left|
s\right\rangle _{1}\left| r\right\rangle _{2}\right) ,\frac{1}{\sqrt{2}}
\left( \left| r\right\rangle _{1}\left| s\right\rangle _{2}+\left|
s\right\rangle _{1}\left| r\right\rangle _{2}\right) \right\} , 
\]

\noindent where $\left| r\right\rangle _{1}$ $\left| r\right\rangle _{2}$
reads particle 1 in site $r$, particle 2 also in site $r$, etc.

The energy eigenvalues $\leftrightarrow $ eigenstates of the Hamiltonian

\[
H_{12}^{\lambda }=\left( 
\begin{array}{llll}
E^{\lambda } & 0 & 0 & 0 \\ 
0 & E^{\lambda } & 0 & 0 \\ 
0 & 0 & 0 & 0 \\ 
0 & 0 & 0 & 0
\end{array}
\right) \text{, with }E^{\lambda }\ \text{discretely above }0\text{, are} 
\]

\[
\ \ 
\begin{tabular}{lll}
$E^{\lambda }$ & $\leftrightarrow $ & $\left| r\right\rangle _{1}\left|
r\right\rangle _{2}$ \\ 
$E^{\lambda }$ & $\leftrightarrow $ & $\left| s\right\rangle _{1}\left|
s\right\rangle _{2}$ \\ 
$0$ & $\leftrightarrow $ & $\frac{1}{\sqrt{2}}\left( \left| r\right\rangle
_{1}\left| s\right\rangle _{2}-\left| s\right\rangle _{1}\left|
r\right\rangle _{2}\right) $ \\ 
$0$ & $\leftrightarrow $ & $\frac{1}{\sqrt{2}}\left( \left| r\right\rangle
_{1}\left| s\right\rangle _{2}+\left| s\right\rangle _{1}\left|
r\right\rangle _{2}\right) .$%
\end{tabular}
\]

\noindent When link $r,s$ has relaxed onto the ground state of $%
H_{12}^{\lambda }$, each link node or qubit is associated with exactly one
particle. This creates the ``lattice'' of qubits $r$ and $s$. For the sake
of simplicity, the two lowest eigenvalues are both $0$; they could as well
be different\footnote{%
which would be necessary if we regarded $H_{12}^{\lambda }$ as acting on
spatial variables.}, provided they are well below $E^{\lambda }$. The link
ground state in $H_{\chi \lambda }=H_{\chi }\otimes H_{\lambda }$ has thus
four degenerate eigenstates (the overall state must be antisymmetrical):

\[
\begin{tabular}{lll}
$0$ & $\leftrightarrow $ & $\frac{1}{\sqrt{2}}\left| 0\right\rangle
_{1}\left| 0\right\rangle _{2}\left( \left| r\right\rangle _{1}\left|
s\right\rangle _{2}-\left| s\right\rangle _{1}\left| r\right\rangle
_{2}\right) $ \\ 
$0$ & $\leftrightarrow $ & $\frac{1}{\sqrt{2}}\left| 1\right\rangle
_{1}\left| 1\right\rangle _{2}\left( \left| r\right\rangle _{1}\left|
s\right\rangle _{2}-\left| s\right\rangle _{1}\left| r\right\rangle
_{2}\right) $ \\ 
$0$ & $\leftrightarrow $ & $\frac{1}{2}\left( \left| 0\right\rangle
_{1}\left| 1\right\rangle _{2}+\left| 1\right\rangle _{1}\left|
0\right\rangle _{2}\right) \left( \left| r\right\rangle _{1}\left|
s\right\rangle _{2}-\left| s\right\rangle _{1}\left| r\right\rangle
_{2}\right) $ \\ 
$0$ & $\leftrightarrow $ & $\frac{1}{2}\left( \left| 0\right\rangle
_{1}\left| 1\right\rangle _{2}-\left| 1\right\rangle _{1}\left|
0\right\rangle _{2}\right) \left( \left| r\right\rangle _{1}\left|
s\right\rangle _{2}+\left| s\right\rangle _{1}\left| r\right\rangle
_{2}\right) $%
\end{tabular}
\]

To avoid triplet states of parallel spins, we introduce the Hamiltonian $%
H_{12}^{\chi }$ operating on the Hilbert space

\[
H_{\chi }=span\left\{ \left| 0\right\rangle _{1}\left| 0\right\rangle
_{2},\left| 1\right\rangle _{1}\left| 1\right\rangle _{2},\frac{1}{\sqrt{2}}%
\left( \left| 0\right\rangle _{1}\left| 1\right\rangle _{2}+\left|
1\right\rangle _{1}\left| 0\right\rangle _{2}\right) ,\frac{1}{\sqrt{2}}
\left( \left| 0\right\rangle _{1}\left| 1\right\rangle _{2}-\left|
1\right\rangle _{1}\left| 0\right\rangle _{2}\right) \right\} , 
\]
where $\left| 0\right\rangle _{1}\left| 0\right\rangle _{2}$ reads: the $%
\sigma _{z}$ component of particle 1 is down, the same for particle 2, etc.

\[
H_{12}^{\chi }=\left( 
\begin{array}{llll}
E^{\chi } & 0 & 0 & 0 \\ 
0 & E^{\chi } & 0 & 0 \\ 
0 & 0 & 0 & 0 \\ 
0 & 0 & 0 & 0
\end{array}
\right) \text{, with }E^{\chi }\text{\ discretely above }0. 
\]

\noindent $H_{12}^{\chi }$ commutes with $H_{12}^{\lambda }$ and lifts the
energy level of the triplet states of parallel spins. The link ground state
remains two-fold degenerate and hosts the spin singlet state $\left| \psi
\right\rangle _{-}$ and the antiparallel spin triplet state $\left| \psi
\right\rangle _{+}$ : 
\begin{equation}
\left| \psi \right\rangle _{\mp }=\frac{1}{2}\left( \left| 0\right\rangle
_{1}\left| 1\right\rangle _{2}\mp \left| 1\right\rangle _{1}\left|
0\right\rangle _{2}\right) \left( \left| r\right\rangle _{1}\left|
s\right\rangle _{2}\pm \left| s\right\rangle _{1}\left| r\right\rangle
_{2}\right) ,
\end{equation}

\noindent upper (lower) signs go together. With simple manipulations, (2)
can be written: 
\begin{equation}
\left| \psi \right\rangle _{\mp }=\frac{1}{\sqrt{2}}\left( \left| \psi
^{^{\prime }}\right\rangle \mp \left| \psi ^{^{\prime \prime }}\right\rangle
\right) ,
\end{equation}
where 
\begin{equation}
\left| \psi ^{\prime }\right\rangle =\frac{1}{\sqrt{2}}\left( \left|
0\right\rangle _{1}\left| r\right\rangle _{1}\left| 1\right\rangle
_{2}\left| s\right\rangle _{2}-\left| 1\right\rangle _{1}\left|
s\right\rangle _{1}\left| 0\right\rangle _{2}\left| r\right\rangle
_{2}\right) =\left| 0\right\rangle _{r}\left| 1\right\rangle _{s},
\end{equation}

\begin{equation}
\left| \psi ^{^{\prime \prime }}\right\rangle =\frac{1}{\sqrt{2}}\left(
\left| 1\right\rangle _{1}\left| r\right\rangle _{1}\left| 0\right\rangle
_{2}\left| s\right\rangle _{2}-\left| 0\right\rangle _{1}\left|
s\right\rangle _{1}\left| 1\right\rangle _{2}\left| r\right\rangle
_{2}\right) =e^{i\delta }\left| 1\right\rangle _{r}\left| 0\right\rangle
_{s}.
\end{equation}
It is readily checked that both terms appearing in $\left| \psi ^{\prime
}\right\rangle $ ($\left| \psi ^{\prime \prime }\right\rangle $) in the
particle representation map onto $\left| 0\right\rangle _{r}\left|
1\right\rangle _{s}$ ($\left| 1\right\rangle _{r}\left| 0\right\rangle _{s}$%
) in qubit representation; $\delta $ is an undefined phase. Thus 
\begin{equation}
\left| \psi \right\rangle _{\mp }=\frac{1}{\sqrt{2}}\left( \left|
0\right\rangle _{r}\left| 1\right\rangle _{s}\mp \left| 1\right\rangle
_{r}\left| 0\right\rangle _{s}\right) ,
\end{equation}
where we have set $\delta =0$ for $\left| \psi \right\rangle $ to behave for
a rotation of the $\chi $ reference in eq. (6) as in eq. (2).

Since $\left| \psi \right\rangle _{-}$ and $\left| \psi \right\rangle _{+}$
are degenerate, we can say as well that the two orthogonal ground
eigenstates of link $r,s$ are $\left| 0\right\rangle _{r}\left|
1\right\rangle _{s}$ and $\left| 1\right\rangle _{r}\left| 0\right\rangle
_{s}$. Therefore the degenerate link ground state has the form:

\begin{equation}
\left| \xi \right\rangle =\alpha \left| 0\right\rangle _{r}\left|
1\right\rangle _{s}+\beta \left| 1\right\rangle _{r}\left| 0\right\rangle
_{s},
\end{equation}
with $\left| \alpha \right| ^{2}+\left| \beta \right| ^{2}=1$.

Let $A_{12}=1-P_{12}$ followed by renormalization be the antisymmetric
operator in particle representation. By definition, the application of $%
A_{12}$ to a generic state filters out and normalizes its antisymmetrical
component. We can see that $A_{12}\left| \psi ^{\prime }\right\rangle
=\left| \psi ^{\prime }\right\rangle $ and $A_{12}\left| \psi ^{^{\prime
\prime }}\right\rangle =\left| \psi ^{^{\prime \prime }}\right\rangle $. In
equivalent terms: $A_{12}\left| 0\right\rangle _{r}\left| 1\right\rangle
_{s}=\left| 0\right\rangle _{r}\left| 1\right\rangle _{s}$ and $A_{12}\left|
1\right\rangle _{r}\left| 0\right\rangle _{s}=\left| 1\right\rangle
_{r}\left| 0\right\rangle _{s}$. Also: $A_{12}\left| 0\right\rangle
_{r}\left| 0\right\rangle _{s}=\left| 0\right\rangle _{r}\left|
0\right\rangle _{s}$ and $A_{12}\left| 1\right\rangle _{r}\left|
1\right\rangle _{s}=\left| 1\right\rangle _{r}\left| 1\right\rangle _{s}$,
if we are dealing with the excited states of energy eigenvalue $E^{\chi }$.
However, such states are forbidden in the link ground state. They should not
be confused with the ``bad'' tensor products: 
\begin{eqnarray*}
\left| 0\right\rangle _{r}\left| 0\right\rangle _{s} &=&\frac{1}{\sqrt{2}}%
\left| 0\right\rangle _{1}\left| 0\right\rangle _{2}\left( \left|
r\right\rangle _{1}\left| s\right\rangle _{2}+\left| s\right\rangle
_{1}\left| r\right\rangle _{2}\right) , \\
\left| 1\right\rangle _{r}\left| 1\right\rangle _{s} &=&\frac{1}{\sqrt{2}}%
\left| 1\right\rangle _{1}\left| 1\right\rangle _{2}\left( \left|
r\right\rangle _{1}\left| s\right\rangle _{2}+\left| s\right\rangle
_{1}\left| r\right\rangle _{2}\right) ,
\end{eqnarray*}
which are symmetrical and violate $A_{12}$. We should note that qubit
notation, and qubit density matrices ($\left| 0\right\rangle
_{r}\left\langle 0\right| _{r}$, etc.), on which the relaxation process
operates, are the same for both kinds of states.

All this becomes, in $H_{rs}=span\left\{ \left| 0\right\rangle _{r}\left|
0\right\rangle _{s},\left| 0\right\rangle _{r}\left| 1\right\rangle
_{s},\left| 1\right\rangle _{r}\left| 0\right\rangle _{s},\left|
1\right\rangle _{r}\left| 1\right\rangle _{s}\right\} $, the projector (or
``symmetry'') $A_{rs}$ defined by the following equations: 
\begin{equation}
A_{rs}\left| \chi \right\rangle _{r}\left| 1-\chi \right\rangle _{s}=\left|
\chi \right\rangle _{r}\left| 1-\chi \right\rangle _{s},\quad A_{rs}\left|
\chi \right\rangle _{r}\left| \chi \right\rangle _{s}=0,
\end{equation}
with $\chi =0,1.$ $A_{rs}$ symmetrization projects a {\em generic} state of $%
H_{rs}$ onto the symmetrical subspace $H_{A_{rs}}$ spanned by the {\em %
legitimate} link states, given by eq. (7). It is understood that the
application of $A_{rs}$ is followed by renormalization. Let $\left\langle
L\right\rangle $ be the expected value of the link energy. Clearly, $A_{rs}$
projection is equivalent to the two mathematically simultaneous conditions: $%
A_{12}$ {\em projection} {\em and} $\left\langle L\right\rangle =0$.

\subsection{A divergence between one and two-way propagation models}

The link state, in the reference where $H_{12}^{\chi }$ is diagonal, is
given by eq. (7). Let us consider continuous transformations of state (7)
which leave $A_{rs}$ symmetry unaltered. Without significant restriction,
state (7) is rewritten: 
\begin{equation}
\left| \xi \right\rangle =\cos \vartheta \left| 0\right\rangle _{r}\left|
1\right\rangle _{s}+\sin \vartheta \left| 1\right\rangle _{r}\left|
0\right\rangle _{s}.
\end{equation}
There is of course a {\em unitary} {\em operator} $R\left( \varphi \right) $%
, acting on the Hilbert space $H_{rs}=span\left\{ \left| 0\right\rangle
_{r}\left| 0\right\rangle _{s},\text{ }\left| 0\right\rangle _{r}\left|
1\right\rangle _{s},\text{ }\left| 1\right\rangle _{r}\left| 0\right\rangle
_{s},\text{ }\left| 1\right\rangle _{r}\left| 1\right\rangle _{s}\right\} $,
which changes $\left| \xi \right\rangle $ of eq. (9) into: 
\begin{equation}
\left| \xi ^{\prime }\right\rangle =\cos \left( \vartheta +\varphi \right)
\left| 0\right\rangle _{r}\left| 1\right\rangle _{s}+\sin \left( \vartheta
+\varphi \right) \left| 1\right\rangle _{r}\left| 0\right\rangle _{s},
\end{equation}
without ever (for all $\varphi $) violating $A_{rs}$ symmetry; here $\varphi
=\omega t$. Since $\vartheta \ $and $\varphi $ will be used to model the
result of gate relaxation, we can assume $\vartheta ,\varphi \geqslant 0$
and $0\leqslant \vartheta +\varphi \leqslant \frac{\pi }{2}$.

It could be seen that $R\left( \varphi \right) $ operates on the overall
link state in an irreducible way. However, for reasons that will become
clear, we are interested in a factorizable operator: $R_{rs}\left( \varphi
\right) =R_{r}\left( \varphi \right) R_{s}\left( \varphi \right) $, where $%
R_{r}\left( \varphi \right) [R_{s}\left( \varphi \right) ]$ operates on the
state of qubit $r$ ($s$), {\em independently} of the state of qubit $s$ ($r$%
). Let $R_{t}\left( \varphi \right) \equiv \left( 
\begin{array}{rr}
\cos \varphi & -\sin \varphi \\ 
\sin \varphi & \cos \varphi
\end{array}
\right) $, with $t=r,s$.

We should note that $R_{rs}\left( \varphi \right) $ does not commute with $%
A_{rs}$. Its application to state (9) introduces the terms $\left|
0\right\rangle _{r}$ $\left| 0\right\rangle _{s}$ and $\left| 1\right\rangle
_{r}\left| 1\right\rangle _{s}$. There are two possible cases:

\begin{enumerate}
\item[(i)]  these terms are the excited states which satisfy $A_{12}$, then
the link goes into a superposition of two energy eigenstates, ground and $%
E^{\chi }$, while its expected energy $\left\langle L\right\rangle $ is
raised, and

\item[(ii)]  these terms are the symmetrical states of parallel spin, which
violate $A_{12}$; such states would be, so to speak, immediately projected
off.
\end{enumerate}

\noindent In this latter case $R_{rs}\left( \varphi \right) $ operates
adiabatically on the link, leaving it in its ground eigenstate, where $%
A_{rs} $ is satisfied\footnote{%
For a study of adiabatic operation on energy eigenstates, see [33].}.
Further below, it will become clear that this is a possible solution.

We shall formalize this, first in the conventional one-way propagation
model. A series of infinitesimal operations $R_{rs}\left( \delta \varphi
\right) $, each followed by $A_{rs}$ symmetrization should be applied until $%
\int \delta \varphi =\varphi $. However this operation suffers from a Zeno
effect which keeps state (9) unaltered: indeed $\lim_{n\rightarrow \infty
}\left[ A_{rs}R_{rs}\left( \frac{\varphi }{n}\right) \right] ^{n}\left| \xi
\right\rangle =\left| \xi \right\rangle $, as it can be checked. In fact, in
first order, $R_{rs}\left( \varphi \right) $ drives the link state out of
the symmetrical subspace $H_{A_{rs}}$ $=$ $span$ $\left\{ \left|
0\right\rangle _{r}\left| 1\right\rangle _{s},\left| 1\right\rangle
_{r}\left| 0\right\rangle _{s}\right\} $ orthogonally to it. $A_{rs}$
projection on $H_{A_{rs}}$ brings the link state back to the starting point.
Therefore $R_{rs}\left( \varphi \right) $ cannot operate adiabatically in a
one-way propagation perspective, or in calculus which is naturally one-way.
It would inevitably lift $\left\langle L\right\rangle $.

However, calculus may not be applicable to a situation which involves a
continuous projection due to $A_{rs}$ symmetrization. As a matter of fact,
during an elementary rotation $\bigtriangleup \varphi =\frac{\varphi }{n}$,
symmetry $A_{rs}$ is continuously violated (if the link remains in its
ground state, the fundamental symmetry $A_{12}$ is violated). This error may
not vanish in the limit $\bigtriangleup \varphi \rightarrow 0$.

A reasonable alternative is to think that rotation and symmetrization should
be applied in a mathematically simultaneous way, so that $A_{rs}$ is never%
{\em \ }violated. This removes the Zeno effect and brings in a two-way
propagation (Section II.F). We consider a {\em free} normalized vector $%
\left| \xi ^{^{\prime \prime }}\right\rangle $ of the Hilbert space $H_{rs}$%
. By this we mean that its amplitudes on $H_{rs}$ basis vectors are free
variables (up to normalization) independent of each other. This vector must
satisfy, for all $\varphi $ or $t\left( \varphi =\omega t\right) ,$ the
following system of mathematically simultaneous equations:

\begin{enumerate}
\item[(i)]  $A_{rs}\left| \xi ^{^{\prime \prime }}\right\rangle =\left| \xi
^{^{\prime \prime }}\right\rangle ,$

\item[(ii)]  $Tr_{s}\left| \xi ^{^{\prime \prime }}\right\rangle
\left\langle \xi ^{^{\prime \prime }}\right| =\cos ^{2}\left( \vartheta
+\varphi \right) \left| 0\right\rangle _{r}\left\langle 0\right| _{r}+\sin
^{2}\left( \vartheta +\varphi \right) \left| 1\right\rangle _{r}\left\langle
1\right| _{r},$

\item[(iii)]  $Tr_{r}\left| \xi ^{^{\prime \prime }}\right\rangle
\left\langle \xi ^{^{\prime \prime }}\right| =\sin ^{2}\left( \vartheta
+\varphi \right) \left| 0\right\rangle _{s}\left\langle 0\right| _{s}+\cos
^{2}\left( \vartheta +\varphi \right) \left| 1\right\rangle _{s}\left\langle
1\right| _{s}.$
\end{enumerate}

Condition (i) ensures that $A_{rs}$ is {\em never} violated and is
equivalent to $A_{12}\left| \xi ^{^{\prime \prime }}\right\rangle =\left|
\xi ^{^{\prime \prime }}\right\rangle $ {\em and} $\left\langle
L\right\rangle =0$. Conditions (ii) and (iii) mean that the density matrix
of each qubit has been rotated by $\varphi $. For all $\varphi $, the above
system of mathematically simulteneous equations yields the solution: 
\[
\left| \xi ^{^{\prime \prime }}\right\rangle =\cos \left( \vartheta +\varphi
\right) \left| 0\right\rangle _{r}\left| 1\right\rangle _{s}+e^{i\delta
}\sin \left( \vartheta +\varphi \right) \left| 1\right\rangle _{r}\left|
0\right\rangle _{s},
\]
as it is readily checked. The further condition that $\left\| \left\langle
\xi \right| \left. \xi ^{^{\prime \prime }}\right\rangle \right\| $ be
maximum (see Section II.F), sets $\delta =0$, thus yielding to state (10).

As we will see (Section II.F), this is exactly an application of a two-way
propagation model\footnote{%
We will use the model developed in [18, 19], whose outline will be given in
Section II.F.} to the projection process induced by $A_{rs}$ (by $A_{12}$
under an adiabatic operation). To sum up, in this scheme, one can operate
separately and adiabatically on the link qubits; the link state undergoes a 
{\em unitary evolution} while remaining in its degenerate ground state, thus
always satisfying $A_{rs}$. This unitary evolution is {\em driven} by
conditions (ii), (iii) and {\em forged} by condition (i), namely continuous $%
A_{rs}$ projection. The fact that the {\em resulting evolution is unitary
thus reversible }(it does not dissipate free energy), is consistent with the
assumption of adiabatic operation.

We will show another way of obtaining the same result. Let qubit $r$ be
connected to the one-qubit Hamiltonian $\left( 
\begin{tabular}{rr}
$E_{r}$ & $0$ \\ 
$0$ & $0$%
\end{tabular}
\right) _{r}$ operating on $span$ $\left\{ \left| 0\right\rangle _{r},\left|
1\right\rangle _{r}\right\} $ and qubit $s$ be connected to the one-qubit
Hamiltonian $\left( 
\begin{tabular}{rr}
$0$ & $0$ \\ 
$0$ & $E_{s}$%
\end{tabular}
\right) _{s}$ operating on $span$ $\left\{ \left| 0\right\rangle _{s},\left|
1\right\rangle _{s}\right\} .$ Disregarding $A_{rs}$, the two qubits would
relax independently of each other onto $\left| 1\right\rangle _{r}\left|
0\right\rangle _{s}$ (a heat bath should provide for both ``shaking'' and
``dumping'').

The expected energy of the two gates in state (9) (the initial condition) is 
$E_{1}=\left( E_{r}+E_{s}\right) \cos ^{2}\vartheta $. Now we set the final
condition that the state of this network simultaneously satisfies $A_{rs}$
and has a lower expected energy, say $E_{2}=\left( E_{r}+E_{s}\right) \cos
^{2}\left( \vartheta +\varphi \right) $. We assume that the energy
difference has gone to the heat bath. This yields to state (10) again, as it
can be checked. Now the unitary evolution from (9) to (10) is {\em driven}
by the independent gate relaxation process, and of course it is forged by
condition (i).

One can think that gate relaxation may not always operate adiabatically on
the link. The excited (antisymmetrical) components $\left| 0\right\rangle
_{r}\left| 0\right\rangle _{s}$ and $\left| 1\right\rangle _{r}\left|
1\right\rangle _{s}$ would appear in a superposition with the link ground
state. However, assuming $E_{r},E_{s}\ll E^{\chi }$, these components would
have a small amplitude and should consequently be short lived, due to the
continuous gate relaxation. The above two-way evolution would still apply to
the Everett worlds where the link is in its symmetric ground eigenstate.

We should further note that, rotating either independent qubit is equivalent
to rotating both; the two rotations are redundant with respect to one
another. This can be checked by removing either condition (ii) or (iii);
just one is sufficient to yield state (10).

We shall now go back to the one-way model. It can be checked that there is
no way of operating separately and adiabatically on the two qubits to
perform a unitary transformation of the link state inside the Hilbert space $%
H_{A_{rs}}$. Gate relaxation will raise the link expected energy $%
\left\langle L\right\rangle $ by some amount, non-infinitesimal this time.
Therefore the link state can undergo a rotation. Then the link would relax
on ground restoring $A_{rs}$ on the rotated state. This would go on
repeatedly. However, this form of quantum annealing works exactly like
classical annealing.

\subsection{Turning Gate Dynamics on}

To sum up, having turned link Hamiltonians on and allowed links to relax
onto ground (independently of each other), the entire lattice of qubits $L$
is obtained, where qubits eigenvalues are submitted to the symmetry
(affecting any pair of qubits connected by a link) $\tilde{A}=\Pi
_{r,s}A_{rs}$. Note that all $A_{rs}$ operate on disjoint Hilbert spaces,
concerning the states of disjoint pairs of nodes, and are pairwise commuting.

After creating $L$ and $\tilde{A}$, gate Hamiltonians are ``turned on'',
then gates are allowed to relax. Here we consider the network at
equilibrium, when all gates and links are in ground state. How this state is
reached will be the subject of Section II.F.

Let $P_{g}$ be the projector on the ground state of gate $\#$ $g$, and $%
\left| \Psi \right\rangle =\sum_{h}\alpha _{h}\left| \Psi _{h}\right\rangle $
the generic network state, where $\left| \Psi _{h}\right\rangle $ is a
tensor product of all qubit eigenstates. At equilibrium, when all gates are
in ground state, for all $P_{g}$ each $\left| \Psi _{h}\right\rangle $
should satisfy $P_{g}\left| \Psi _{h}\right\rangle =\left| \Psi
_{h}\right\rangle .$ This yields to $\left| \Psi _{h}\right\rangle $s which
satisfy all gates. Each $\left| \Psi _{h}\right\rangle $ should also satisfy 
$\tilde{A}\left| \Psi _{h}\right\rangle =\left| \Psi _{h}\right\rangle $.
This yields to$\left| \Psi _{h}\right\rangle $s which satisfy all links. All 
$P_{g}$ operate on disjoint sets of qubits, thus on disjoint Hilbert spaces,
and are pairwise commuting. Let $P=\Pi _{g}P_{g}$, where $g$ runs over all
gate labels. The foregoing simultaneous equations are summarized by:

\begin{equation}
\tilde{A}P\left| \Psi \right\rangle =\left| \Psi \right\rangle ,
\end{equation}

\noindent Any $\left| \Psi _{h}\right\rangle $ satisfying eq. (11) satisfies
all gates and links and is thus a solution (for the case with no solution,
see Section II.A). Substituting $\tilde{A}P$ with $P\tilde{A}$ yields to the
same conclusion. Even if $\tilde{A}$ and $P$ did not commute, the network
ground states are eigenstates of both. However, the propagator of the actual
relaxation process is shaped by $\tilde{A}$ and commutes with it.

This approach would suffer no interaction with the environment provided $L$
and $\tilde{A}$ remain in place: entanglement with a reservoir initially in
state $\left| R\right\rangle $ changes each $\left| \Psi _{h}\right\rangle
\left| R\right\rangle $ into $\left| \Psi _{h}\right\rangle \left|
R_{h}\right\rangle $, where $\left| R_{h}\right\rangle $ is now correlated
with $\left| \Psi _{h}\right\rangle $, without altering the selection
performed by eq. (11).

\subsection{Relaxation Time}

We shall now discuss relaxation time. Let us consider a simple ``network''
of just one link $r,s$ and one qubit gate, whose Hamiltonian operates on $%
span$ $\left\{ \left| 0\right\rangle _{s},\left| 1\right\rangle _{s}\right\} 
$ and has the form $\left( 
\begin{tabular}{ll}
$E_{s}$ & $0$ \\ 
$0$ & $0$%
\end{tabular}
\right) _{s}$. Let the initial state of this network be the symmetrical
state $\left| \Psi \left( t_{1}\right) \right\rangle =\left| 0\right\rangle
_{r}\left| 1\right\rangle _{s}$. For the sake of simplicity, only qubit $r$
is assumed to be in direct interaction with a heat bath. At equilibrium, the
(symmetrical) ground state should instead be $\left| 1\right\rangle
_{r}\left| 0\right\rangle _{s}$.

We shall apply the one-way propagation model first. We will limit ourselves
to examining the network initial evolution from $\left| 0\right\rangle
_{r}\left| 1\right\rangle _{s}$. This is readily done and provides useful
insight. Since the initial state is factorized, relaxation of qubit $s$ and
heat bath perturbation on qubit $r$ are (initially and in first
approximation) independent of each other.

The initial independent relaxation of gate s can be modeled by the two-step
evolution:

\begin{equation}
\left| \Psi \left( t_{1}\right) \right\rangle _{s}=\left| 1\right\rangle
_{s}\rightarrow \left| \Psi \left( t_{2}\right) \right\rangle _{s}=\cos
\triangle \varphi \left| 1\right\rangle _{s}+e^{i\delta }\sin \triangle
\varphi \left| 0\right\rangle _{s},
\end{equation}

\noindent \noindent where $t_{2}>t_{1}$, and $\delta $ is an undefined phase
due to entanglement with the outside world. We are using the method of
random phases: the average over $\delta $ of the density matrix
corresponding to the state on the right of the arrow, yields $\cos
^{2}\triangle \varphi \left| 1\right\rangle _{s}\left\langle 1\right|
_{s}+\sin ^{2}\triangle \varphi \left| 0\right\rangle _{s}\left\langle
0\right| _{s}$, as expected.

The initial independent heat bath perturbation on qubit $r$ can be modeled
by the two-step evolution: 
\begin{equation}
\left| \psi \left( t_{1}\right) \right\rangle _{r}=\left| 0\right\rangle
_{r}\rightarrow \left| \psi \left( t_{2}\right) \right\rangle _{r}=\cos
\triangle \vartheta \left| 0\right\rangle _{r}+e^{i\delta ^{\prime }}\sin
\triangle \vartheta \left| 1\right\rangle _{r},
\end{equation}
where $\triangle \vartheta $ is a random phase.

We can approximate the network initial evolution by symmetrizing the tensor
product of the two qubits: 
\begin{eqnarray}
\left| \psi \left( t_{1}\right) \right\rangle &=&\left| 0\right\rangle
_{r}\left| 1\right\rangle _{s}\rightarrow  \nonumber \\
\left| \psi \left( t_{2}\right) \right\rangle &=&\cos \triangle \varphi \cos
\triangle \vartheta \left| 0\right\rangle _{r}\left| 1\right\rangle
_{s}+e^{i\delta ^{^{\prime \prime }}}\sin \triangle \varphi \sin \triangle
\vartheta \left| 1\right\rangle _{r}\left| 0\right\rangle _{s},
\end{eqnarray}
up to normalization; here $\delta ^{^{\prime \prime }}=\delta +\delta
^{^{\prime }}$.

This is an interesting result. If $\triangle \vartheta =0$, the evolution
remains{\em \ frozen }in its initial state $\left| 0\right\rangle _{r}\left|
1\right\rangle _{s}$ (see eq. 14). The relaxation of gate $s$ is conditioned
by the action of the heat bath on qubit $r$. This means that, so to speak,
by a lucky coincidence the network can relax, otherwise it cannot. Likely,
this would mean frustration in a more complex network.\ In this
approximation, one can see no reason why quantum annealing should perform
better than classical annealing.

Before applying the two-way propagation model ($TW$) developed in [18, 19],
we shall outline it. This model was meant to justify the projection
undergone by a quantum state during various measurement operations.

Both forward and backward propagations are referred to the same direction of
time in the episystem, on the basis of the following argument. Time can be
seen as a {\em parameter} in the endosystem. It is the {\em direction} {\em %
of} {\em causality} in the endosystem $-$ either forward{\em \ }or backward
in time $-$ that makes the difference between the two propagations.
Therefore, both propagations are represented by {\em kets evolving forward
in time}: one associated with {\em causality forward in time}, the other
with {\em causality backward in time}. Because of the absence of a
privileged direction of causality, the two propagations should be
indistinguishable.

Indistinguishability requires that (i) each propagation undergoes the same
transformations of the conventional one-way propagation, {\em up to an
overall phase, which is irrelevant to each propagation}, and (ii) the
relative phase of the two propagations is either $0$ or $\pi $.

As a consequence of indistinguishability, the two propagations should
proceed in coherent superposition. This superposition is the {\em complete
description} of the propagation. Both initial and final conditions,
comprising normalization, should be applied to it, whereby the initial (or
indifferently) final condition of each propagation is just a free
un-normalized state (completely undefined before the application of all
conditions) of the system Hilbert space.

State projection (due to either a measurement or a watchdog effect) implies
a $\pi $ shift in the relative phase of the two propagations. Consequently,
their superposition changes from constructive to destructive interference or
vice-versa. Clearly, this can justify the discontinuous evolution of the
quantum state throughout a projection.

The foregoing model yields to the following simple rules to be applied to
the superposition (the complete description):

\begin{enumerate}
\item[(i)]  the state before reduction $\left| \psi \left( t_{1}\right)
\right\rangle $ coincides with the usual state;

\item[(ii)]  the state after reduction $\left| \psi \left( t_{2}\right)
\right\rangle $ is a free normalized state of the system Hilbert space,
subject to the final conditions and to the supplementary condition that

\item[(iii)]  $\left\| \left\langle \psi \left( t_{1}\right) \right| \left.
\psi \left( t_{2}\right) \right\rangle \right\| $ is maximum.
\end{enumerate}

A preparation, or a propagation thereof, can be the state before projection.
The final conditions can be set from the future, for example they can be the
result of partial or complete measurement, unforeseeable from the past.

As it should be clear now, in Section II.D we have applied the two-way
propagation model schematized above. The result was that $A_{rs}$ works like
interference (destructive due to projection, constructive due to
renormalization) on the relaxation of independent gates.

We shall now apply the {\em TW}\ model to the simple network introduced at
the beginning of this Section.

The independent relaxation of gate $s$ is modeled by the two-step evolution
(12). The gate energy in $\left| \psi \left( t_{2}\right) \right\rangle _{s}$
is $E_{s}\cos ^{2}\triangle \varphi $ (from now on, by energy in a state we
will mean the expected value of the energy operator in that state).

The heat bath perturbation on qubit $r$ is modeled by (13). Let the operator 
$\left( 
\begin{tabular}{rr}
$E_{r}$ & $0$ \\ 
$0$ & $-E_{r}$%
\end{tabular}
\right) _{r}$ represent the energy of the qubit-heat bath interaction. Its
expected value in $\left| \psi \left( t_{2}\right) \right\rangle _{r}$ is $%
E_{r}\cos 2\triangle \vartheta $. $\triangle \vartheta $ can be assumed to
be a random variable with uniform distribution in $\left[ 0,2\pi \right] $,
consequently $\cos 2\triangle \vartheta $ is zero on the average. We further
assume $E_{r}\ll E_{s}$.

Since the gate and the heat bath operate independently (given that $A_{rs}$
works as interference), the network energy at time $t_{2}$ is $E=E_{s}\cos
^{2}\triangle \varphi +E_{r}\cos 2\triangle \vartheta $.

The network state at time $t_{2}$ is a free normalized state of the network
Hilbert space, submitted to the {\em final conditions}:

\begin{enumerate}
\item[(i)]  $\left| \Psi \left( t_{2}\right) \right\rangle $ satisfies $%
A_{rs}$;

\item[(ii)]  the energy in $\left| \psi \left( t_{2}\right) \right\rangle $
is $E$.
\end{enumerate}

This yields the two-step evolution

\begin{equation}
\left| \psi \left( t_{1}\right) \right\rangle =\left| 0\right\rangle
_{r}\left| 1\right\rangle _{s}\rightarrow \left| \psi \left( t_{2}\right)
\right\rangle =\cos \Delta \varphi ^{^{\prime }}\left| 0\right\rangle
_{r}\left| 1\right\rangle _{s}+e^{i\delta ^{^{\prime \prime }}}\sin \Delta
\varphi ^{^{\prime }}\left| 1\right\rangle _{r}\left| 0\right\rangle _{s},
\end{equation}
where $\triangle \varphi ^{\prime }$ is still to be determined. The former
condition (iii) is irrelevant here.

The energy in $\left| \psi \left( t_{2}\right) \right\rangle $ is $E_{s}\cos
^{2}\triangle \varphi ^{\prime }+E_{r}\cos 2\triangle \varphi ^{\prime }$.
Therefore $\triangle \varphi ^{^{\prime }}$ must be a solution of the
equation 
\[
E_{s}\cos ^{2}\triangle \varphi ^{\prime }+E_{r}\cos 2\triangle \varphi
^{^{\prime }}=E_{s}\cos ^{2}\triangle \varphi +E_{r}\cos 2\triangle
\vartheta . 
\]
It can be readily checked that there is a unique solution in the interval of
interest. If $E_{r}\ll E_{s}$, $\triangle \varphi ^{\prime }$ $-$ i.e.
network relaxation $-$ closely follows the independent gate relaxation $%
\triangle \varphi $. Moreover, $\triangle \varphi =\triangle \varphi
^{\prime }$ on the average.

This is quite different from the one-way model. Since the network energy is
the sum of the energies of the {\em independent} network elements,
frustration is never met. For example, the absence of a heat bath ($%
\triangle \vartheta $ or $E_{r}=0$) does not freeze relaxation, unlike in
the one-way model.

This can be generalized to any Boolean network. Condition (iii), moving in
time together with continuous projection, allocates the overall energy among
the network degrees of freedom. Still, the overall energy decreases as if
all network elements relaxed independently of each other and frustration is
never met. Although a dynamic model is still lacking, this qualitative
difference should yield an essential advantage over classical simulated
annealing.

Let us note that evolution (15) can represent an identical rotation of the
two independent qubits $r$ and $s,$ obtained by rotating just qubit $s$. We
must put $E_{r}=0$ (thus obtaining $\triangle \varphi ^{\prime }=\triangle
\varphi $) and $\delta ^{^{\prime \prime }}=0$, which ensures rotation
additivity; this is legitimate since we are now dealing with a pure state
rotation. We find again the result of Section II.D.

\medskip

\section{Discussion}

We have developed a form of quantum annealing computation where network
wiring is implemented by a quantum symmetry $\tilde{A}$ related to particle
statistics. How this quantum wiring operates can be interpreted in two
different ways.

In a one-way, conventional, propagation model, this wiring does not exhibit
any special feature with respect to classical wiring. Disregarding
tunnelling, this quantum model does not yield any improvement over classical
annealing.

In an unconventional $-$ two-way $-$ propagation model, particle statistics
works as a projection. $\tilde{A}$ would continuously {\em project} the
state of the independently relaxing gates on a ``symmetric'' subspace where
network wiring is satisfied. This is naturally a blunt conjecture $-$ this
work should be considered an exploration of the possible role played by
particle statistics (viewed as projection) in quantum computation.

We should remark that something similar appears in other situations
involving particle statistics. For example, given two {\em non-interacting}
identical spin $\frac{1}{2}$ fermions whose spatial wave functions overlap,
there is a {\em correlation} between their mutual distance and the character
of their spin state, either singlet or triplet. This correlation, due to
antisymmetrization, creates a sort of ``wiring'' between two otherwise
independent parts.

Although an implementation model is still lacking, we see no reason to think
that this is out of reach. In the example where $\lambda $ labels the sites
of a spatial lattice and $\chi $ labels the $\sigma _{z}$ component of the
spin $\frac{1}{2}$ particle hosted by each site, the link behaviour is
likely to be related to exchange interaction.

Were this model implementable $-$ for example in the case of the single link
and a one qubit Hamiltonian (Section II.F) $-$ relaxation speed could be
tested. In principle, this should decide which propagation model $-$ either
one or two-way $-$ applies.

The hypothesis of a computation speed-up based on chronology violation links
this work with D. Deutsch's paper on computation along closed time-like
lines. From that paper we quote: ``it is curious that the analysis of a
physical situation which might well not occur should yield so many insights
in quantum theory. But that is the nature of thought experiments. Perhaps we
should also bear in mind that a frequently observed effect of time is to
convert thought experiments into real ones''. Placing chronology violation
within the time-energy uncertainty relation and the projection of an
evolving state induced by particle statistics, might be a step towards
feasibility.

This research has been supported by Elsag Bailey, a Finmeccanica company.
Thanks are due to A. Ekert, D. Finkelstein, L. Levitin, S. Lloyd, D. Monti,
M. Rasetti and T. Toffoli for the useful suggestions.

\end{document}